\documentclass[journal]{IEEEtran}

\usepackage{color}

\usepackage{cite}

\ifCLASSINFOpdf
\usepackage[pdftex]{graphicx}
\else
\fi

\usepackage{amsmath}
\usepackage{amsfonts}

\usepackage[ruled,vlined]{algorithm2e}

\usepackage{algorithmic}

\usepackage{multirow}
\usepackage{authblk}

\begin{document}

	\title{Top-level Design and Simulated Performance of the First Portable CT-MR scanner}
	\author[a,$\ast$]{Yuting Peng}
    \author[b,$\ast$]{Mengzhou Li}
    \author[a]{Jace Grandinetti}
    \author[b,$\ddag$]{Ge Wang}
    \author[a,$\ddag$]{Xun Jia}
    \affil[a]{Innovative Technologies Of Radiotherapy Computations and Hardware (iTORCH) Laboratory, Department of Radiation Oncology, University of Texas Southwestern Medical Center, Dallas, TX, 75390, USA.}
    \affil[b]{Biomedical Imaging Center/Cluster, Rensselaer Polytechnic Institute, Troy, 12180, NY, USA.}
    \affil[$\ddag$]{To whom correspondence should be addressed: {email:Xun.Jia@UTSouthwestern.edu}, {email:wangg6@rpi.edu}}
    \affil[$\ast$]{The first two authors contributed equally.}

	\markboth{Top-level Design and Simulated Performance of the First Portable CT-MR scanner}{}

	\maketitle


\begin{abstract}
Multi-modality imaging hardware can be integrated in a single gantry to collect diverse datasets for complementary information and spatiotemporal correlation, with excellent examples including PET-CT and PET-MRI. However, there is no CT-MRI prototype up to today due to technical challenges and associated cost-benefit considerations. Thanks to the rapid development of medical imaging, it becomes feasible now to integrate cost-effective CT and MRI imagers together for portability, popularity, and point of care. In this paper, we present the top-level design of the first portable CT-MRI system and evaluate its imaging performance via realistic numerical simulations. In this CT-MRI system, the magnet made of two NdFeB rings of about 40.0 cm radius forms a magnetic field of about 57 mT at the isocenter and has a gap of 11.3 cm to accommodate the rotating CT gantry. 
The targeted MR imaging field of view (FOV) is a sphere of $\sim$15 cm in diameter and that of CT is approximately 20 cm diameter in axial direction and 5 cm in longitudinal direction. Our results show a great potential of such a hybrid system. The proposed CT-MRI system will be valuable in applications such as imaging in underdeveloped countries, disaster scenes and battle fields.

\end{abstract}

	\begin{IEEEkeywords}
		Computed tomography (CT), magnetic resonance imaging (MRI), multi-modality imaging
	\end{IEEEkeywords}

	\ifCLASSOPTIONpeerreview
	\fi
\IEEEpeerreviewmaketitle

\section{Introduction}\label{Introduction}

\IEEEPARstart{M}{}ultimodality medical imaging is well known to deliver tremendous utilities in a wide spectrum of clinical scenarios\cite{townsend2008multimodality}. In fact, each imaging modality produces information in a well-defined scope, as characterized by spatial/temporal resolution, signal-to-noise ratio, structural, functional and molecular features, etc. Although these well-defined scopes may overlap, no single modality could derive all the information another modality offers for a major clinical task, as modern medicine often calls for comprehensive evaluation of the subject from a variety of aspects with different imaging modalities. While it is possible to acquire images of different modalities on separate scanners, thereby achieving multi-modal imaging via post-scan image fusion, this approach inevitably suffers from issues such as required image registration, associated uncertainty, not to mention challenges associated with logistics, burdens and complications of long scan time between scanners caused by moving patients and dynamic changes in them, especially in contrast-enhanced studies and critical care conditions where synchrony and efficiency of imaging procedures is vital. 

Over the years, several types of hybrid scanners have been developed for simultaneous multi-modal imaging. PET-CT is now widely available, serving as an indispensable tool for cancer diagnosis, staging, and treatment response assessment \cite{blodgett2007pet}. Novel PET-MRI scanners allow characterization of metabolic activities empowered by MRI to delivery structural information in rich soft tissue contrast without ionizing radiation \cite{judenhofer2008simultaneous}. More importantly, in these situations, imaging data from different modalities acquired at the same time of the patient permit joint image reconstruction and processing to synergistically take advantages of different images, enhancing image quality and performance in clinical tasks. In contrast to these multi-modality imaging progresses, CT-MRI is the only major remaining multi-modality imaging technology yet to be developed. In our previous Vision 20/20 paper\cite{wang2015vision}, we have suggested that such a CT-MRI combination could have profound impacts on several major clinical scenarios such as cardiac diagnosis and contrast-enhanced cancer imaging.

Nonetheless, integrating CT and MRI scanners for simultaneous imaging is a rather challenging undertaking. Conventional MRI scanners are often designed with a high-strength (commonly, 1-3 T) homogeneous (of the order of a few parts per million) magnetic field for a high signal to noise ratio (SNR) and imaging performance within a clinically acceptable scan time. This means that the field is susceptible to perturbations from nearby ferromagnetic materials and the imaging performance is easily affected by the x-ray tube and detector assembly, when being integrated with a CT scanner. The weak MR signal is also prone to radiofrequency (RF) signals generated by nearby devices, which deteriorates SNR of MR images. Conversely, the fringe magnetic field over the x-ray tube and detector assembly affects their functions due to deflection of the electron beam and malperformance of electronics in the magnetic field. Last but not the least, bulky MR and CT components have to be integrated in a compact space, further enhancing the challenge of the CT-MRI system design. 

Fortunately, over the past years there have been exciting developments towards addressing these challenges. To combine x-ray radiography with MRI scanner, researchers at Stanford group demonstrated the feasibility of operating an x-ray tube and a detector in the magnetic field \cite{fahrig2001truly, fahrig2005performance}. In the radiotherapy field, recent advances of MRI-guided radiotherapy achieved by a combination of MRI with medical linear accelerator demonstrated clinical MR imaging functions in close proximity to a powerful x-ray production device \cite{lagendijk2014magnetic,raaymakers2017first}. 

In this study, we report our on-going effort towards prototyping the first portable CT-MRI system employing a non-conventional MR scanner design. Specifically, we propose an ultra-low-field (ULF) MR scanner operating at a ~57 mT flux density range together with an inhomogeneous magnetic field design. The medical relevance of ultra-low-field MRI has recently been demonstrated in several examples \cite{marques2019low,mcdaniel2019mr,cooley2021portable,liu2021low}. We prefer an inhomogeneous field design to reduce the technical challenges associated with building a highly homogeneous magnet. The feasibility of using an inhomogeneous magnetic field for MRI has been previously shown in several systems, such as single-sided MRI scanners. In our design, the low strength and inhomogeneous field make the MRI system more robust to perturbations from the CT components, and the low-fringe field facilitates the integration with the CT subsystem. 

\section{Results}
\subsection{Top-level system design}
\begin{figure}[htbp]
    \centering
    \includegraphics[width=0.5\textwidth]{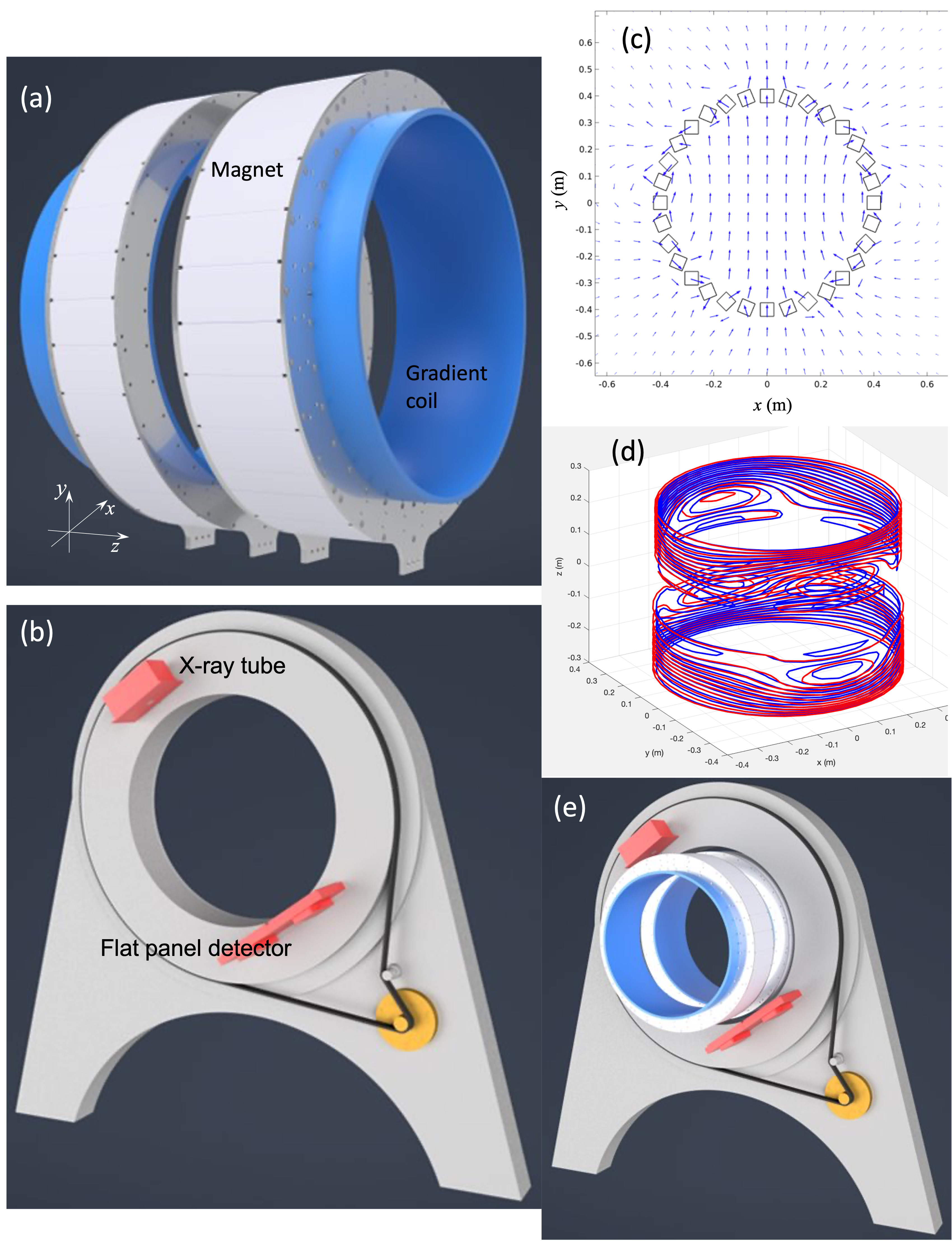}
    \caption{Top-level design of the CT-MRI system. (a) MRI subsystem design (gradient coils inside of magnets shown in blue. RF coil is not drawn); (b) CT subsystem design; (c) Orientation of one Halbach ring and magnetic flux density field shown at the $z=0$ plane; (d) y-gradient coil design with blue showing primary layer and red showing shielding layer; and (e) an overall system rendering.}
    \label{fig:design}
\end{figure}

\subsubsection{MRI subsystem}
The overall design of the CT-MRI scanner is shown in Figure \ref{fig:design}. We consider and ULF MR design for the purpose of cost reduction and easy intergration with CT subsystem. For this ULF MRI subsystem (Figure \ref{fig:design}(a)), The magnet consists of two rings of NdFeB permanent magnets with a gap of 11.33 cm for the x-ray beam of the CT subsystem to pass through. Both rings are configured to orient the permanent magnets in a Halbach format to form the magnetic field fairly uniformly in the y direction. On one ring, 34 N52 bar magnets of $5.08\times 5.08\times 15.24$ cm$^3$ are aligned along a circle of radius $40.6$ cm, while the second ring consists of 32 bar magnets of $5.08\times 5.08\times 20.32$ cm$^3$ on a circle of $40.0$ cm radius. These magnets, their sizes, and arrangment define a magnetic field of about $57$ mT at the isocenter. The Halbach ring and magnetic flux density distribution on the $z=0$ plan is shown in Figure \ref{fig:design}(c). The magnetic field is slightly stronger towards the smaller ring, naturally inducing a field gradient which can be utilized for spacial encoding along this direction (z axis) to reduce the required number of gradient coils. The resulting gradient strength is about $25$ mT/m. To hold the magnet bars in place, 3D printing technology can be used to create molds based on the designed positions and orientations of magnet bars. Aluminum plates will be used to secure the molds in place. The inner radii the magnet assembly are 36.8 cm and 36.1 cm for the large and small rings respectively.

Two sets of self-shielded gradient coils are designed for spacial encoding along patient lateral directions (x and y directions). The four layers of the two gradient coils are parallel to the inner surface of the magnet, occupying a 2.5 cm space (Figure \ref{fig:design}(a)). The target field method is employed to determine the coil patterns \cite{turner1993gradient} to generate a linearly varying field within the planned field of view (FOV) of 15 cm in diameter, while maximally minimizing magnetic field at nearby metal parts to avoid eddy current effect. Figure \ref{fig:design}(d) presents the resulting coil pattern for the y-gradient coil. Note that there is a $11.5$ cm gap at the center of each coil to avoid blocking the x-ray beam of the CT system.

We plan to use a solenoid transmit/receive RF coil. It follows a standard design for an MR scanner and is not shown here. As for the data acquisition pipeline, standard MRI techniques can be used with a spectrometer for sequence control, RF and gradient amplifiers for signal amplification. Care must be taken to accommodate relatively low SNR due to the low field strength and signal loss caused by dephasing in the presence of the inhomongeneous field.

\subsubsection{CT subsystem}
A cone-beam scanning geometry is adopted for the CT subsystem, as shown in Figure \ref{fig:design}(b). The source-to-detector distance is 1,125 mm, and source-to-isocenter distance 500 mm. These values are close to those for a medical CT scanner and large enough to avoid geometrical interference with the MRI subsystem. The medical pulse x-ray source of interest has a focal spot of $\sim$0.5~mm, a half-cone angle of $18^\circ$, a kVp range between 100 to 400 kVp, and a current range between 1 to 13 mA. To cover a major part of the field of view, we consdier two PaxScan 2520DX flat panel detectors (VAREX Imaging, Salt Lake City, UT) combined to form a larger detector. The x-ray sensor material is amorphous silicon and equipped with a direct deposit CsI conversion screen. The detector has a pixel pitch of 127 $\mu \mathrm{m}$ and supports a resolution 1 lp/mm at $>48\%$. The detector array is of $1536\times 1920$ pixels. The energy detection range is 40 to 160 keV. The intrinsic frame rate is 12.5/s but can be increased when working in binning modes. The detector weighs $\sim$ 5.5l bs. We consider that the detector working in the $4\times4$ binning mode. Due to the obstruction of the MRI assembly, the x-ray beam is confined to a half cone-beam angle $3^\circ$ in the longitudinal direction to avoid scattering from the MRI components. In this setting, the effective detector array becomes $384\times 960$ pixels with pixel pitch of 508 $\mu \mathrm{m}$. With this setting, the CT FOV is approximately 20 cm diameter in axial direction and 5 cm in longitudinal direction.

The x-ray source and detector are mounted on a rotation gantry and a slip ring design is used for continuous data communication with the rotor, as well as power delivery to the tube and panel. This is a drum type slip ring with a inner diameter of 950 mm and supports a maximum rotation speed of 120 rpm.

\subsection{Magnetic field distribution}

\begin{figure}[t]
\centering
\includegraphics[width=0.5\textwidth]{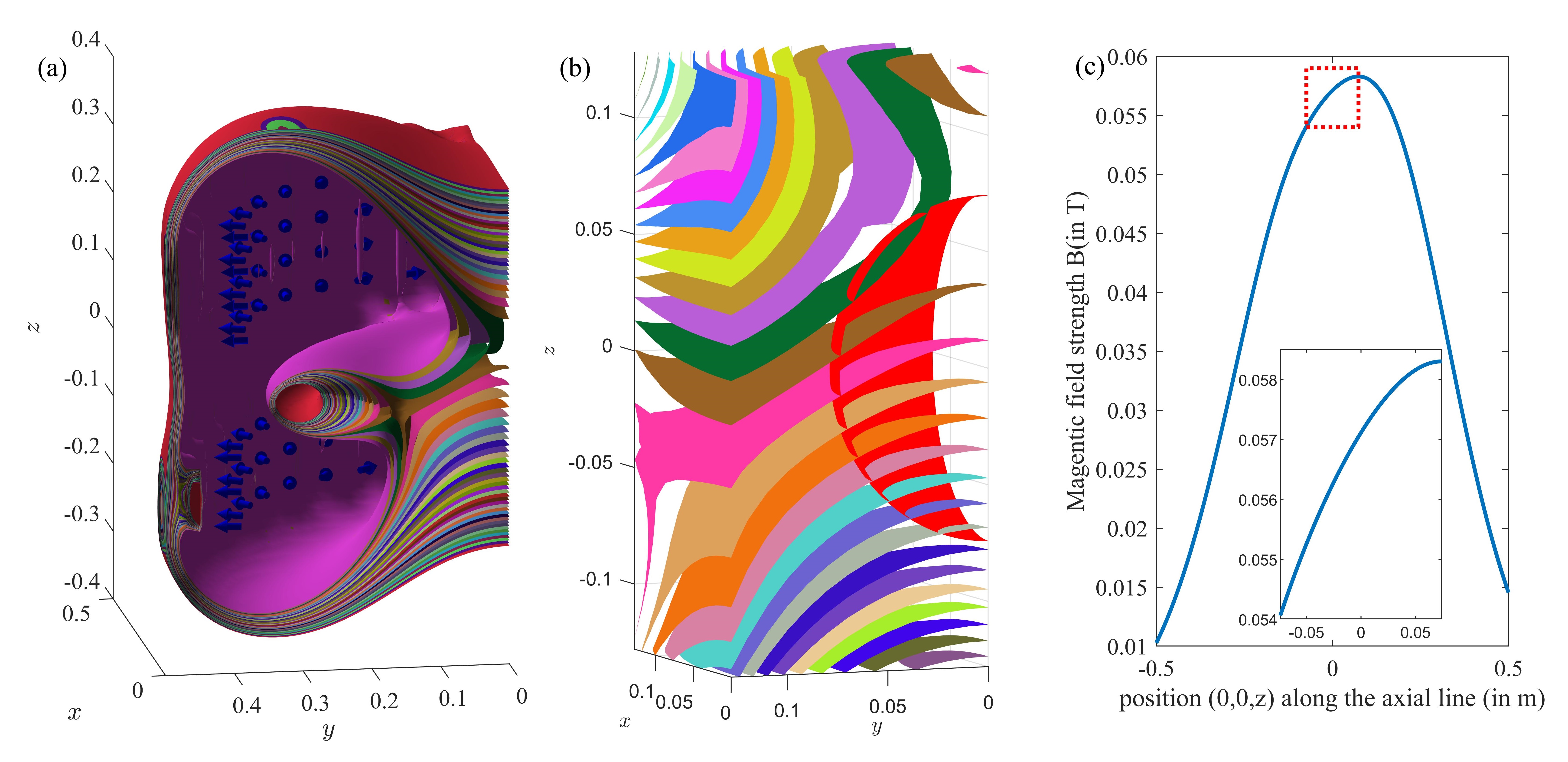}
\caption{(a) and (b) Iso-surfaces of various flux density strengths and the zoom-in view around the iso-center; and (c) the magnetic flux density strength along the z axis through the iso-center. The blue arrows in (a) indicates the magnetic moment direction of each NdFeB magnetic cube. The red sphere indicates the targeted FOV.}
\label{fig:Magneticfield} 
\end{figure}

Figure \ref{fig:Magneticfield} presents the magnetic flux density distribution generated by the main magnet. For a targeted spherical imaging field of view (FOV) of 15 cm in diameter, the flux density strength is 55~68 mT. As shown in Figure \ref{fig:Magneticfield}(c), the magnetic flux density generally decays monotonically along the z direction, creating a gradient about 25 mT/m in the FOV, which offers the opportunity for spatial encoding along this direction without a gradient coil. In Figure \ref{fig:Magneticfield}(a) and (b), the iso-surfaces of various magnetic filed strengths are plotted. These surfaces are planned for data acquisition and image reconstruction by performing 2D scans on these surfaces. The separation between neighboring surfaces is $\sim$ 0.4 cm. It is clearly shown that these surfaces are not planar, which hence requires image domain interpolation to generate a volumetric image defined on the Cartesian grid.

The magnetic flux density strength drops to $\sim$ 20 mT outside the magnet at a radius of 50 cm from iso-center, for the positions of the x-ray tube, and $\sim$ 3 mT at the CT detector position with a radium of 62.5 cm from the iso-center. These low magnetic field is expected to be not interfering the normal functions of these CT components \cite{wen2005robust}.

\subsection{Imaging performance}

\begin{figure}[t]
\centering
\includegraphics[width=0.5\textwidth]{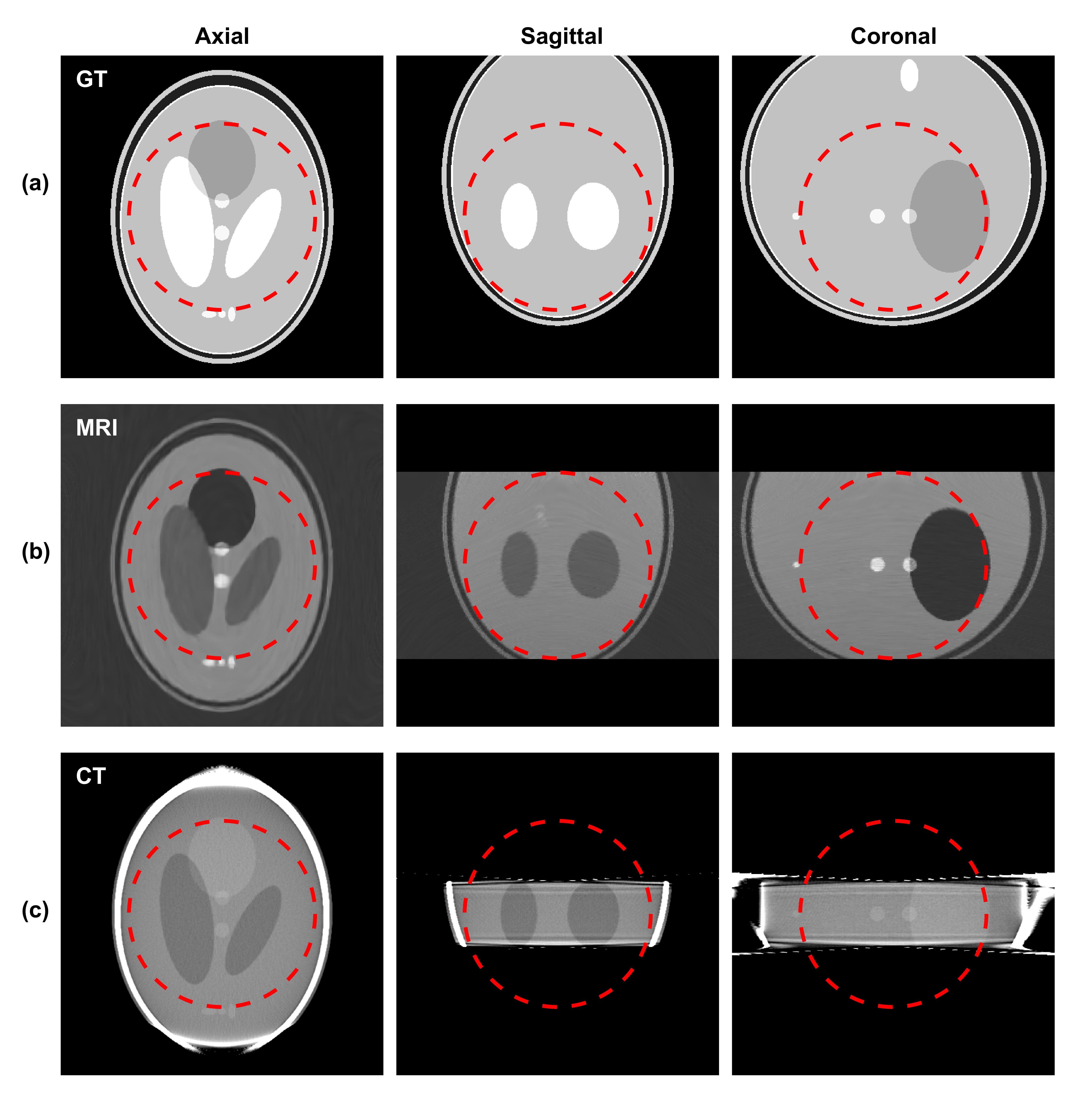}
\caption{Orthogonal views of (a) the proton density of the 3D Shepp-logan phantom, (b) simulated MR images (TR=1,000 ms,TE=80 ms, and NEX=16), and (c) simulated CT images. The dashed red circles indicate the targeted MR FOV of 15 cm in diameter.}
\label{fig:CT_MRI_SL} 
\end{figure}

We first present simulation results for the Shepp-Logan phantom in Figure~\ref{fig:CT_MRI_SL}. For the MR simulation, the relaxation time TR and TE are 1,000 ms and 80 ms, respectively. The slice thickness is chosen as 4mm, and the voxel size on the axial planes is $1 \times 1$ mm$^2$. To reduce noise, we repeated data acquisition with the number of excitations (NEX) being set to 16. The images in Figure~\ref{fig:CT_MRI_SL}(a) display proton density of the phantom in three orthogonal views to show the structure of the phantom. Figure~\ref{fig:CT_MRI_SL}(b) shows the resulting MR image. The CT simulation results of the Shepp-Logan phantom are in Figure~\ref{fig:CT_MRI_SL}(c). The image voxel size is $0.5$ mm$^3$. Due to the narrow x-ray beam through the gap between the two magnet rings, the longitudinal coverage of the CT scan is $\sim 5$ cm. 

\begin{figure}[htbp]
    \centering
    \includegraphics[width=\linewidth]{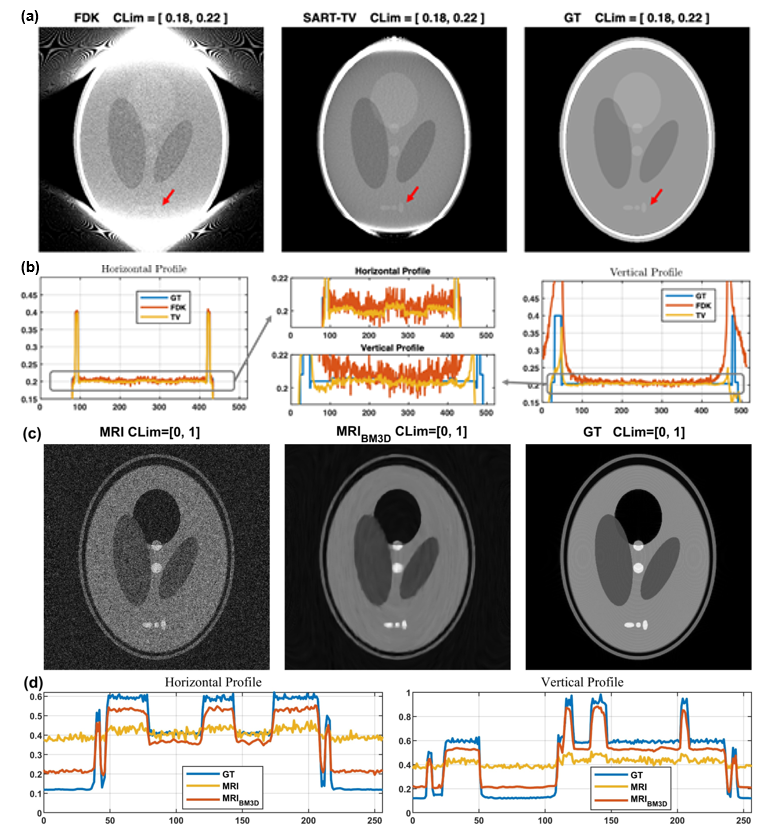}
    \caption{(a, b) FDK and SART-TV reconstructions and (c, d) MRI images without and with BM3D denoising of the Shepp-Logan phantom against the ground truth (GT) at the central axial slice shown in Figure~\ref{fig:CT_MRI_SL}. (b) Horizontal and vertical profiles of CT reconstruction result through the center of the images displayed in (a). The unit for the vertical axis in the plots is cm$^{-1}$ and the image display window in (a) is [0.18, 0.22] cm$^{-1}$. (d) Horizontal and vertical profiles of MRI reconstructions result through the center of the images displayed in (c). Image display window in (d) is [0 1].}
    \label{fig:PhantomImg}
\end{figure}

Figure~\ref{fig:PhantomImg} presents the axial view of the phantom in CT and MR images with different reconstruction and processing methods. For CT, on the lateral direction, the limited FOV introduces truncation artifacts in the conventional FDK reconstruction result, including the bright distorted parts in the image and the cupping effect in the vertical profile. Also, the three small structures close to the FOV boundary can be hardly discerned from the background due to the low contrast between the tumor and the gray matter. Those issues are successfully addressed with the interior tomography technique as demonstrated in the SART-TV result ($\lambda=0.0003$). The tumors with only 10 HU contrast from the background (pointed by the red arrow) are clearly revealed thanks to the noise suppression capability of the technique. The cupping and truncation artifacts are also successfully removed from the reconstruction, and the profiles align well with the ground truth counterparts.

As for the MR images, the image reconstructed by Fourier algorithm had a large amount of noise due to the low magnetic field strength and field inhomogeneity-induced signal dephasing. Using the BM3D algorithm effectively suppressed image noises.

\begin{figure}[t]
\centering
\includegraphics[width=0.5\textwidth]{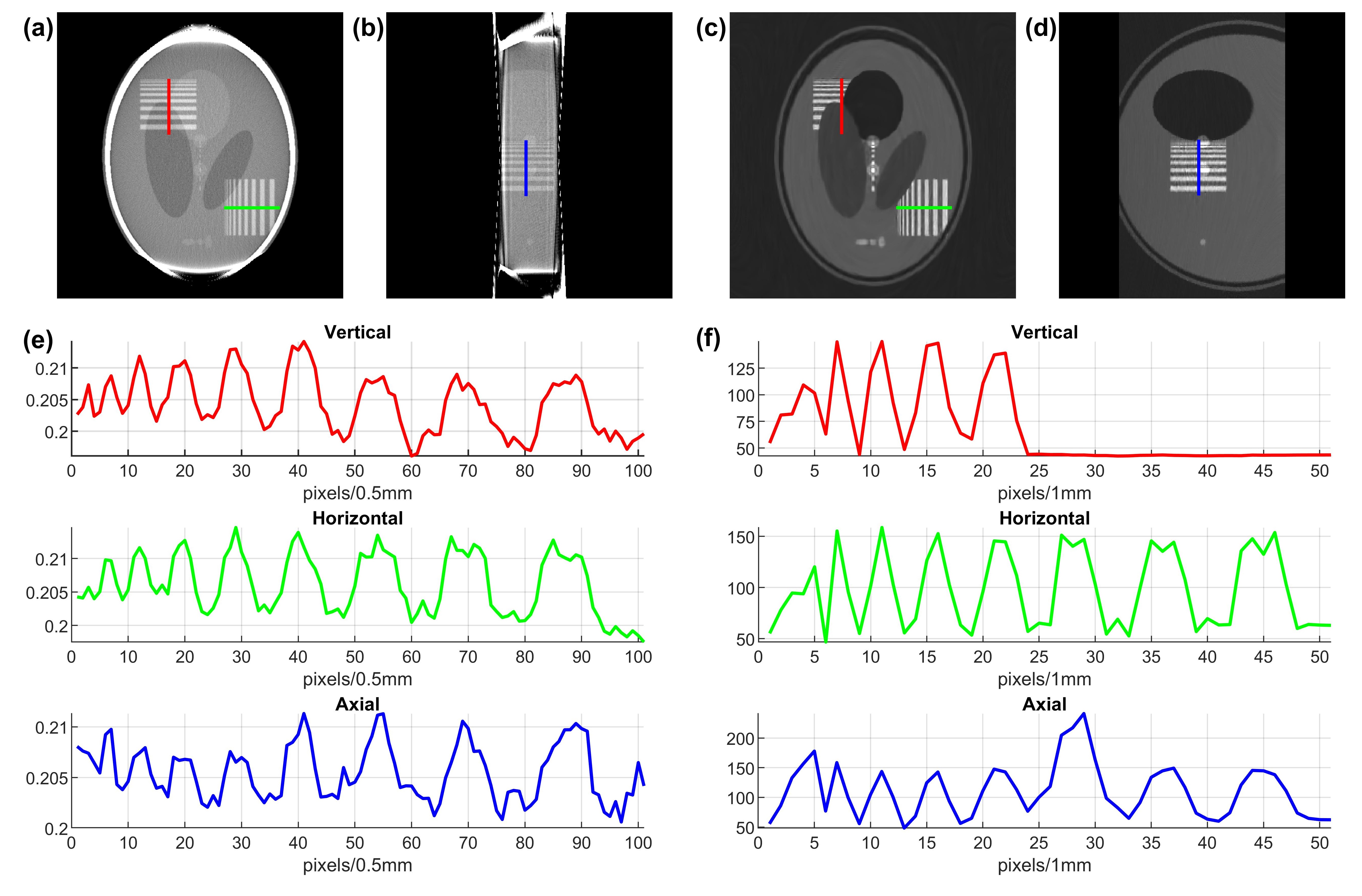}
\caption{\label{fig:CT_MRI_barPattern} CT (a-b) and MR (c-d) images showing bar patterns embedded in the Shepp-Logan phantom. (e)-(f) Line profiles along the bar patterns for three orthogonal directions in CT (e) and MR (f) images. }
\end{figure}

Figure~\ref{fig:CT_MRI_barPattern} presents the results to demonstrate image resolution performance of the imaging systems. As such, we embed in the phantom bar patterns along the three major directions, with the bar width ranging from 0.5 mm to 5 mm. It is observed that the smallest bar at width 0.5 mm can be still observed, although the contrast reduces for small bars.

\begin{figure}[htbp]
\centering
\includegraphics[width=0.5\textwidth]{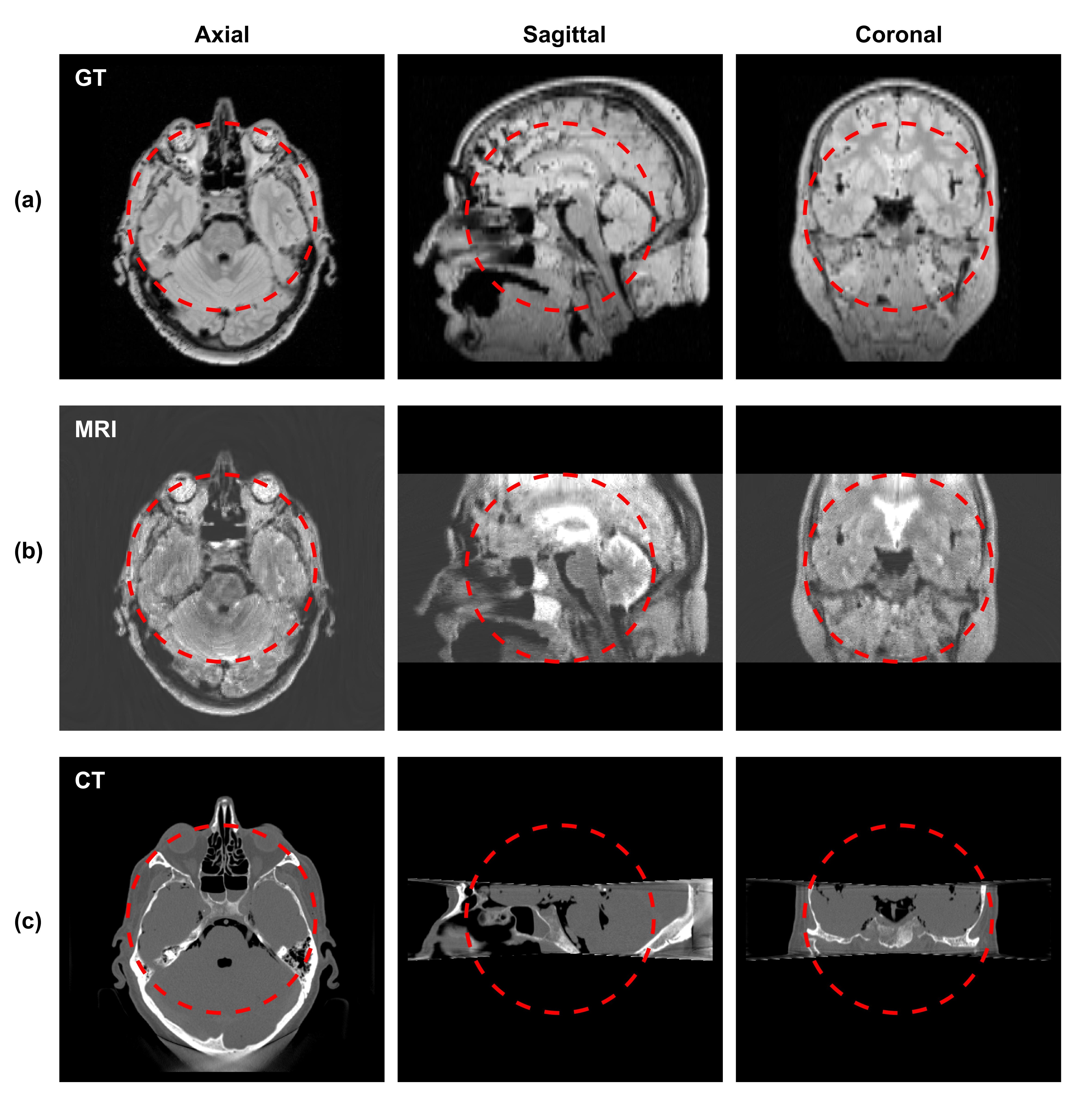}
\caption{Orthogonal views of (a) the spin density of the VHP brain phantom, (b) simulated MR images (TR=1,000 ms,TE=80 ms, and NEX=16), and (c) simulated CT images. The dashed red circles indicate the MR FOV of 15 cm in diameter.}
\label{fig:CT_MRI_head} 
\end{figure}

\begin{figure}[htbp]
\centering
\includegraphics[width=0.5\textwidth]{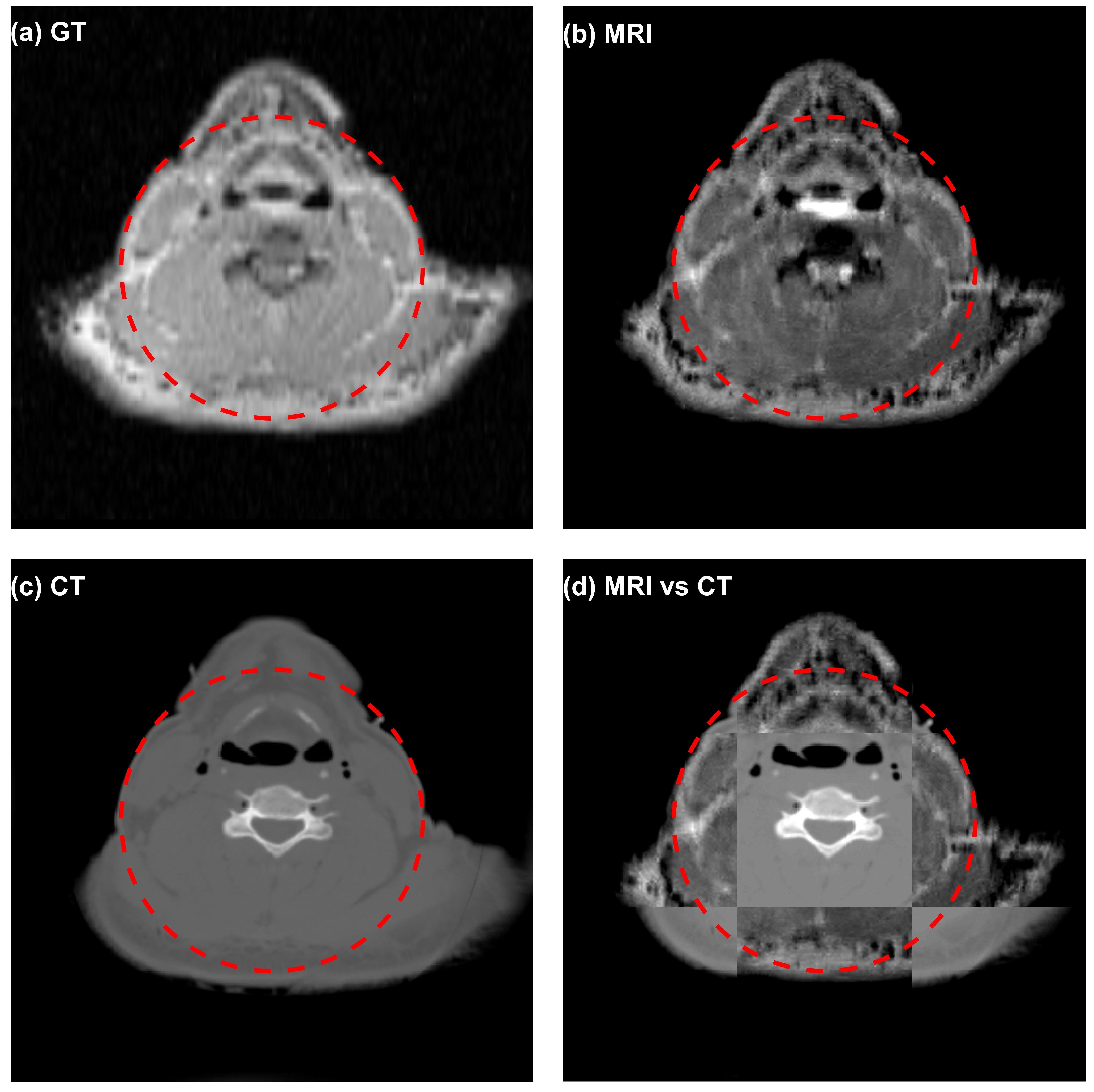}
\caption{\label{fig:Neck_AXIAL} Axial planes of (a) proton density image of the VHP neck phantom, (b) simulated MR image (TR=1,000 ms,TE=80 ms, and NEX=16), (c) simulated CT image, and (d) blended view of MRI and CT images. The dashed red circles indicate the targeted FOV of 15 cm in diameter.}
\end{figure}

\begin{figure}[htbp]
\centering
\includegraphics[width=0.5\textwidth]{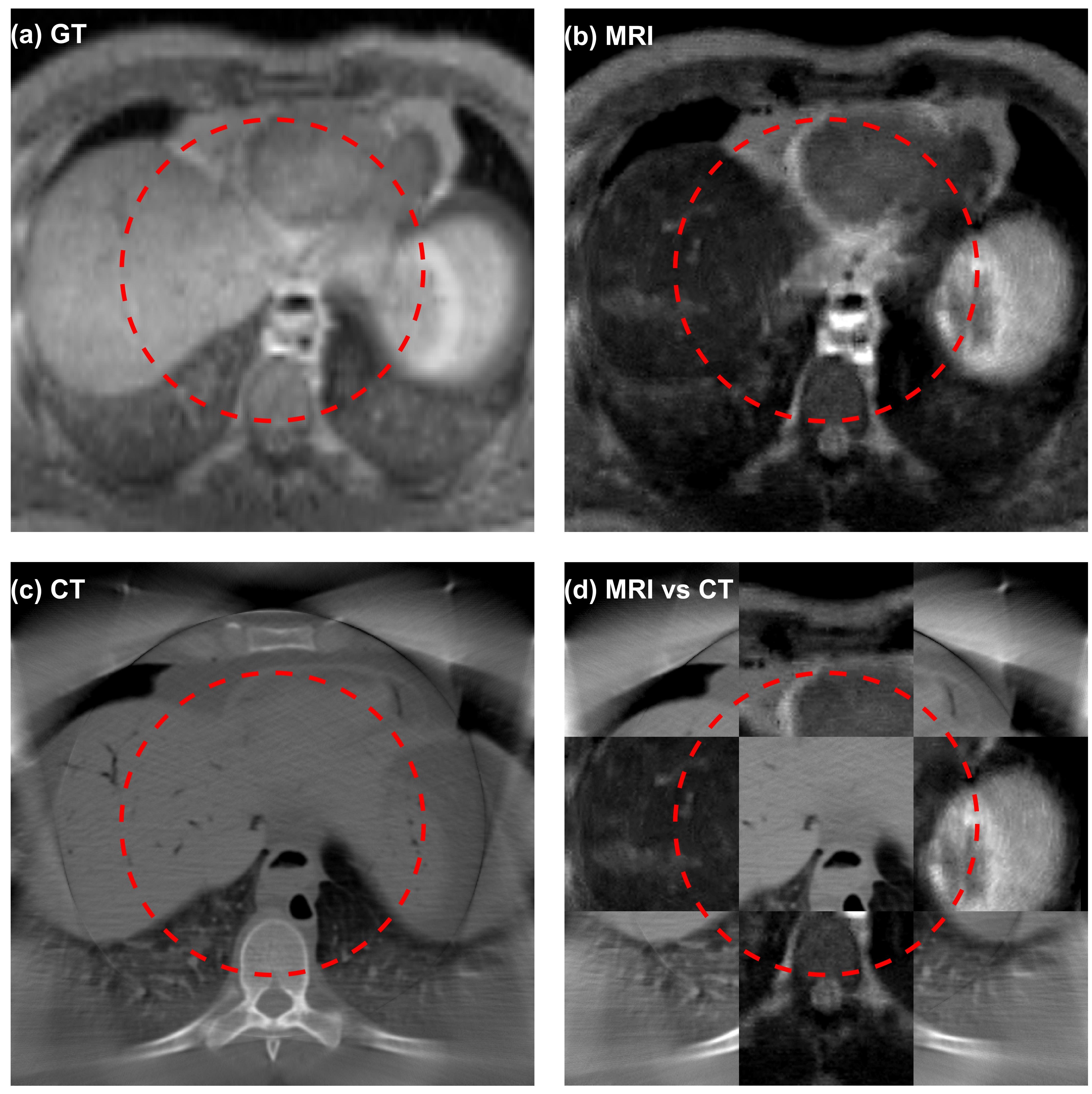}
\caption{\label{fig:Chest_AXIAL} Axial planes of (a) proton density image of the VHP chest phantom, (b) simulated MR image (TR=1,000 ms,TE=80 ms, and NEX=16), (c) simulated CT image, and (d) blended view of MRI and CT images. The dashed red circles indicate the targeted FOV of 15 cm in diameter.}
\end{figure}

To illustrate the performance of the proposed system with more clinical relevance, we present in Figure~\ref{fig:CT_MRI_head}-\ref{fig:Chest_AXIAL} results for different body sites using the VHP data. In the brain case shown in Figure~\ref{fig:CT_MRI_head}, we remark that the relatively low resolution along superior-inferior direction in the MR image is intrinsically caused by the phantom itself. In fact, in the VHP MR image dataset the slides spacing is 3 mm, which limited the image resolution in the simulation results. 

We present in Figures~\ref{fig:Neck_AXIAL} and \ref{fig:Chest_AXIAL} the results from the neck and chest sites. In each figure, in addition to presenting the simulation results of the CT and MR images, we also generated a blended view by displaying CT and MRI images in different squares. The capability of integrated CT-MRI imaging in the same spatial coordinate system is expected to offer advantages in a variety of clinical tasks, such as disease diagnosis and therapy planning.

\section{Discussions}

In this paper, we have proposed a top-level design for an integrated CT-MRI system. The design is featured by an unconventional MRI subsystem with an ultra-low inhomogeneous main magnetic field and a slim CT subsystem. The advantages of taking this approach are multiple folds. The first and most profound advantage is to ensure system integration and compatibility between CT and MR components. Reducing the strength of the main magnetic field makes the field at CT components, i.e. x-ray tube and flat panels, work without any significant electromagnetic interference. If additional shielding is needed to reduce the field strength further, it is straight forward to use shielding materials, such as mu metals, to isolate the CT and MRI subsystems bi-directionally. 

Employing an inhomogeneous field design naturally increased tolerability to uncertainties in the system engineering, as compared to the traditional homogeneous field design. Realizing the homogeneous field usually requires complex approaches, such as active and passive shimming, to achieve the targeted homogeneity level of a few parts per million. In contrast, with the inhomogeneous field, our design does not require a specific level of homogeneity, and the imaging and reconstruction process is performed with the magnetic field distribution as prior information. This conceptual change is expected to substantially reduce the engineering challenge in the design of the magnetic field distribution. In the actual hardware system, even if the achieved field deviated from the design, as long as the deviation is small, e.g. still satisfying the condition of monotonicity along the z-axis, the field is still usable for MRI. 

As demonstrated in our study, the magnetic main field varying along the z-axis can naturally serve as the spatial encoding field along this direction, hence eliminating the need for one of the three spatial gradient coils in classical MR scanners. This helps reduce the complexity of the system integration within a compact space, lowering the system cost and simplifying the system maintenance. 

Our design is also featured by a relatively large bore size of $67\sim 68$ cm in diameter at the inferior and the superior sides. This is larger than commonly used MR scanners. This large bore allows flexible patient positioning to place the anatomical part of interest in the imaging FOV. In principle, a smaller bore size would help increase field strength, being beneficial from SNR perspective. For this ultra-low-field design, a slight increase in field strength due to reduced bore diameter would not significantly translate to a significant SNR improvement. Hence, we elected a large bore size design to give freedom of patient positioning. This design should be favorable for patients with claustrophobia.

The current design aims at a low-cost solution for a CT-MRI system. The magnet was uniquely designed for this system, which uses $>200$ Nd cubic magnets arranged in two rings with a gap to accommodate a CT subsystem. Together with aluminum plates and other supporting structures, the cost of the magnet is $\sim \$20$k. The MR gradient coils, RF coils, and spectrometer follow the standard design for conventional MRI, and price is expected to be $\sim \$50$k. Together with  the CT subsystem with an expected hardware cost of $\sim\$100$k, the total cost of the proposed system is estimated to be under $\sim\$250$k including other accessories, which is very attractive in a good number of use cases. These numbers are bulk estimates and may subject to variation. If the system is for commercial production, additional factors have to be included, such as development costs. We also note that we chose a permanent magnet design for cost consideration in this on-going research project. Increasing field strength using superconducting magnet is certainly possible, although at an increased cost and complexity of the magnet design to ensure compatibility with CT. On the other hand, it is also possible to generate a suitable low-strength magnetic field using conducting coils.

The current study has several limitations. First and foremost, being a simulation study, it can only demonstrate the performance of the proposed system roughly, whereas the overall feasibility and performance will inevitably require actual construction of the system. Several aspects of the system have not been included in the current study, including CT-induced RF interference to MRI, mechanical stability of CT rotation and calibration, etc. We are in the process of building the ultra-low-field MR scanner as a standalone system to first evaluate the MRI performance. The prototype MRI subsystem will also serve as a basis for the evaluation of the compatibility with the CT subsystem. Down the road, we will move forward towards the complete portable CT-MRI system construction, demonstration and translation. 

Second, the novel design in this study also means unique challenges. Historically, ultra-low-field MRI has been neglected due to low SNR. In recent years, interest in this direction has been renewed to build MRI scanners for its favorable features such as low cost and high mobility. Recent advances in image processing, especially deep learning-based imaging \cite{wang2020deep,shen2020introduction}, have offered a potential solution to overcome the low SNR challenge in the ultra-low-field regime, ensuring their values in clinical applications. In our study, imaging performance was reported using the Fourier transform-based reconstruction method together with a denoising step. To increase SNR, we employed a scab protocol with a relatively large number of repetitions, hence causing the long data acquisition time. Applying advanced deep reconstruction and processing techniques to improve image quality is beyond the scope of this paper, and will be our future work. We expect that it will be highly rewarding to employ deep learning techniques to mitigate noise and together with super-resolution techniques to maintain image resolution. 

Another major direction to improve the proposed system performance is joint CT-MRI reconstruction. With the proposed system, CT and MR images are naturally aligned in the same coordinate system, sharing many similar features, such as edges. Taking advantages of this explicit correlation, many implicit corrective relationships will synergize CT and MRI reconstructions. Previous studies demonstrated feasibility and potential advantages of this joint reconstruction approach over conventional image domain regularization\cite{knoll2016joint}. With recent advances in deep learning, it should be possible and desirable training a unified model to learn a joint distributions of CT and MR images, which can serve as a strong prior-information to guide the reconstruction process and improve the resultant image quality.

Another limitation of the proposed system is related to the inhomogeneous magnetic field. Compared to conventional homogeneous field MRI systems, the inhomogeneous field causes intra-voxel dephasing and hence signal loss. We need to use spin-echo MR sequences for refocusing dephased spins during data acquisition. Nonetheless, this inevitably increases data acquisition time compared to rapid gradient based sequences. While some sequences, such as turbo spin echo, are expected to accelerate data acquisition, the inherent signal loss, coupled with the already low SNR in this low-field setup, calls for advanced image reconstruction and processing techniques to optimize the image quality. 

Comparing to MRI, the CT subsystem basically follows a standard CT scanner design. Because of the relative maturity of the CT technology, we expect less hurdles in developing the CT subsystem. Again, simulation studies were used as a proof-of-principle study to establish the feasibility of our overall idea. More thorough investigation will be performed in the next step to analyze all factors that affect the imaging performance. For example, Monte Carlo radiation transport simulation may be used to evaluate the x-ray scatter effect.

\section{Methods}

\subsection{Simulation of MR data acquisition and image reconstruction}

We consider 2D spin-echo type sequences for MR data acquisition, because it helps to alleviate signal dephasing caused by the main magnetic field inhomogeneity. Without loss of generality, for each 2D acquisition, a RF wave of frequency $f$ only excites spins around a curved iso-magnetic field surface defined by the condition $f=\gamma B_0(x,y,z)$, where $B_0(x,y,z)$ is the magnetic flux density field, and $\gamma$ is the gyromagnetic ratio. For a given phantom defined by volumetric images of longitudinal relaxation time $T_1(x,y,z)$,  transverse relaxation time $T_2(x,y,z)$, and spin density $\rho(x,y,z)$, let a spin-echo sequence be specified with repetition time $TR$ and echo time $TE$, the integrated signal for a pulse with a slice selection frequency $f$ is 
\begin{equation}\label{eq:pareto mle2}
\begin{split}
S(k_x,k_y) &=\sum_{x,y,z}I_w[|f-\gamma B_0(x,y,z)|]\rho(x,y,z)\\
&\left[1-\exp(-\frac{TR}{T_1(x,y,z)})\right]\exp[-\frac{TE}{T_2(x,y,z)}]\\
&\exp[-TE{(\delta \gamma \Delta B)}]\exp[-i(k_xx+k_yy)],
\end{split}
\end{equation}
where $I_w[\cdot]$ is an index function with $I_w[x]=1$ if $|x|<w/2$ and 0 otherwise. $w$ is the bandwidth of the RF pulse. The term $\exp[-TE{(\delta \gamma \Delta B_0)}]$ describes dephasing effect due to the magnetic field gradient, and $\delta$ is a constant obtained by calibrating this model using a MR scanner in our lab.

Noise cannot be neglected in an ULF MRI scan due to the low magnetic field yielding a weak signal and the field inhomogeneity induced dephasing effect. Hence, we add to $S(k_x,k_y)$ white noise with standard deviation $\sigma=C_0 \sqrt{B_0}/\sqrt{T_{ACQ}}$ \cite{macovski1996noise}, where $C_0$ is a calibration constant obtained empirically on an MRI scanner available in our lab, and $T_{ACQ}$ is the total acquisition time.

\begin{figure}[t]
\centering
\includegraphics[width=0.49\textwidth]{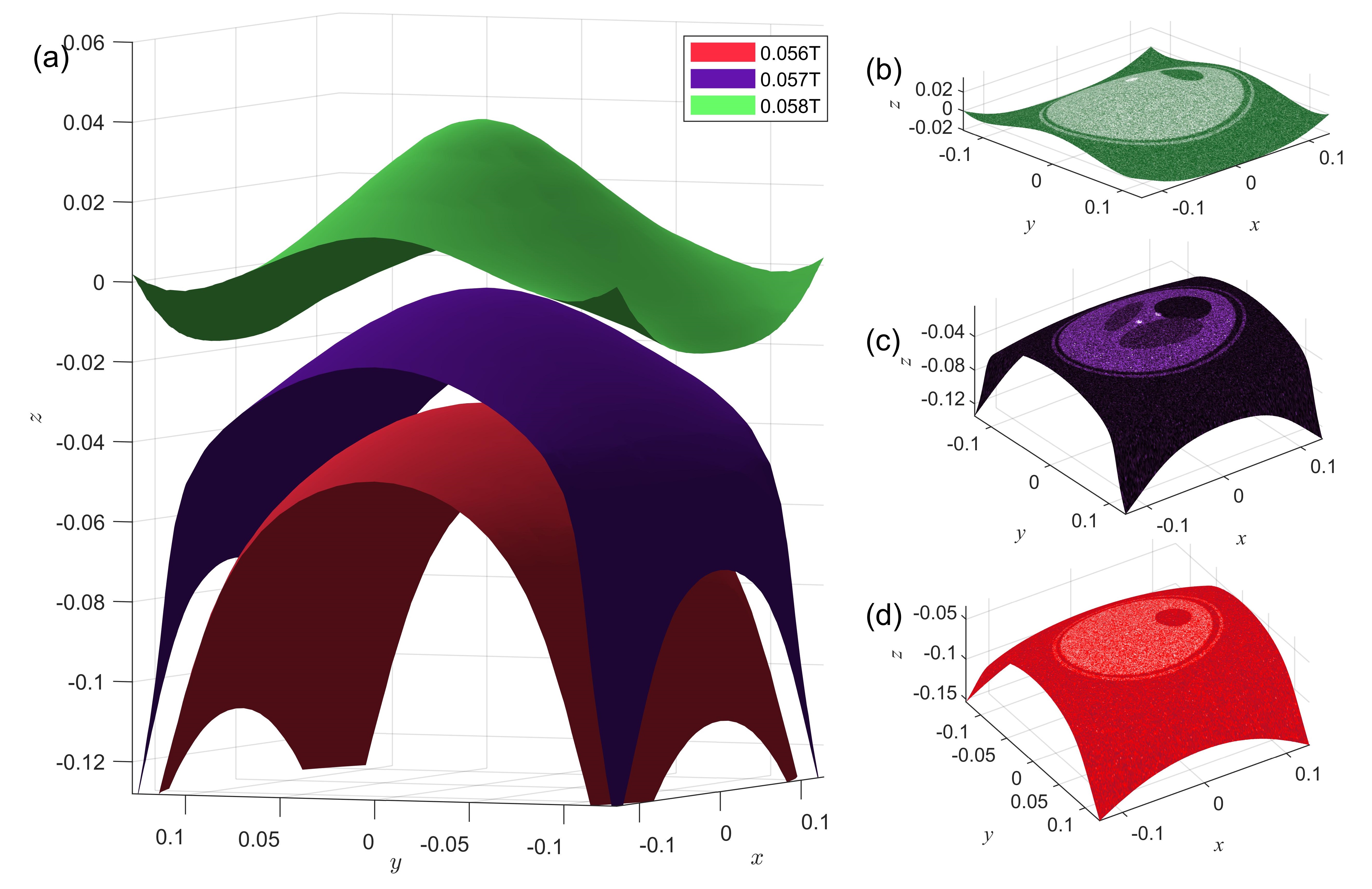}
\caption{(a) Iso-surfaces of three magnetic flux density strengths; and (b-d) 2D images obtained by MR scans that are defined on these iso-surfaces.}
\label{fig:mapping} 
\end{figure}

After collecting $k$ space data, a 2D image $m(x,y)$ is reconstructed using Fourier-transform based reconstruction algorithm. To suppress image noise, we denoise the images using BM3D algorithm \cite{maggioni2012nonlocal}. After collecting images for a series of RF frequency, we can obtain 2D images defined on a set of curved iso-surfaces of magnetic flux density, each characterized by $f=\gamma B_0(x,y,z)$, as illustrated in Figure \ref{fig:mapping}. To generate a volumetric image, we resample these 2D images to a Cartisian grid according to the known positions of these iso-surfaces.

\subsection{Simulation of CT data acquisition and image reconstruction}

To simulate the CT imaging process, we consider a circular scan mode with 720 projection views per rotation. To simulate x-ray projections, a monochromatic radiation model is utilized for simplicity, and the detected signal is expressed as
\begin{equation}
    p = \mathbf{f}_{Poisson}\left[N_0\exp(-P{\mu}(x,y,z))\right], 
    \label{eq:CTdet}
\end{equation}
where $P$ stands for the CT projection system matrix, ${\mu}(x,y,z)$ the volumetric image of the x-ray attenuation coefficient, $\mathbf{f}_{Poisson}[\cdot]$ the Poisson process, ${p}$ the recorded projections in a vector form, and $N_0$ the incoming number of photons, which is set to $3\times10^6$ in this study.

With the simulated x-ray data, an image is reconstructed using the standard Fredkamp-Davis-Kress algorithm \cite{feldkamp1984practical}. Since the CT FOV is relatively small compared to the whole human body, even with the combination of two detectors, the CT system is subject to lateral truncation of projections. Hence, an iterative method for interior tomography \cite{yu2009compressed} is used to improve reconstruction results. Specifically, a Simultaneous algebraic reconstruction technique with a total variation (SART-TV) regularization is coded to solve the optimization problem
\begin{equation}
    \mathrm{argmin}_{\mu} \left\|P{\mu}(x,y,z) + \log({{p}}/{N_0})\right\|^2 + \lambda \|\nabla {\mu}(x,y,z)\|_1,
\end{equation}
where $\lambda$ is a parameter balancing the data fidelity term and the sparsity constraint. The ASTRA toolbox is used during the forward projection and volumetric reconstruction \cite{van2015astra}.

\subsection{Performance evaluation}

We  performed numerical simulations with two phantoms to demonstrate the performance of the proposed CT-MRI scanner. The first phantom is a 3D Shepp-Logan phantom. This phantom represents typical brain tissues including scalp, bone, cerebrospinal fluid (CSF), gray and white matters, and tumor. For CT simulation, the attenuation values are chosen in reference to the original specification of the 2D Shepp-Logan phantom \cite{shepp1974fourier}. As the Shepp-Logan phantom only defines relative density values to water in the interval [1.0, 2.0], we assign x-ray attenuation based on the density values. For MR simulation, the parameters of proton density $\rho$, relaxition times $T_1$ and $T_2$ are set in reference to \cite{gach20082d}.

To bear more clinical relevance, we employ another phantom generated from the publicly-available, anatomically detailed, 3D CT and MRI database from the Visible Human Project (VHP) of the National Library of Medicine \cite{ackerman1998visible}. The male patient images are chosen in this study. X-ray attenuation coefficients are derived from the CT images. In the MRI simulation, since only proton density weighted, $T_1$ weighted and $T_2$ weighed MR images are provided in the database, we derive the proton density image, longitudinal and transverse relaxation times by solving signal equations associated with the MR sequences.




\section{Acknowledgements}
This work is supported in part by funds from the National Science Foundation (NSF: \# 1636933 and \# 1920920) and Cancer Prevention and Research Institute of Texas (\# RP200573).

\section{Author contributions statement}
Y.P. and M.L. performed computational studies. J.G. conducted studies on MR system design. All authors contributed to the research idea and wrote and reviewed the manuscript.




\bibliographystyle{IEEEtran}
\bibliography{reference}

\begin{thebibliography}{10}
\providecommand{\url}[1]{#1}
\csname url@samestyle\endcsname
\providecommand{\newblock}{\relax}
\providecommand{\bibinfo}[2]{#2}
\providecommand{\BIBentrySTDinterwordspacing}{\spaceskip=0pt\relax}
\providecommand{\BIBentryALTinterwordstretchfactor}{4}
\providecommand{\BIBentryALTinterwordspacing}{\spaceskip=\fontdimen2\font plus
\BIBentryALTinterwordstretchfactor\fontdimen3\font minus
  \fontdimen4\font\relax}
\providecommand{\BIBforeignlanguage}[2]{{%
\expandafter\ifx\csname l@#1\endcsname\relax
\typeout{** WARNING: IEEEtran.bst: No hyphenation pattern has been}%
\typeout{** loaded for the language `#1'. Using the pattern for}%
\typeout{** the default language instead.}%
\else
\language=\csname l@#1\endcsname
\fi
#2}}
\providecommand{\BIBdecl}{\relax}
\BIBdecl

\bibitem{townsend2008multimodality}
D.~Townsend, ``Multimodality imaging of structure and function,'' \emph{Physics
  in Medicine \& Biology}, vol.~53, no.~4, p.~R1, 2008.

\bibitem{blodgett2007pet}
T.~M. Blodgett, C.~C. Meltzer, and D.~W. Townsend, ``Pet/ct: form and
  function,'' \emph{Radiology}, vol. 242, no.~2, pp. 360--385, 2007.

\bibitem{judenhofer2008simultaneous}
M.~S. Judenhofer, H.~F. Wehrl, D.~F. Newport, C.~Catana, S.~B. Siegel,
  M.~Becker, A.~Thielscher, M.~Kneilling, M.~P. Lichy, M.~Eichner
  \emph{et~al.}, ``Simultaneous pet-mri: a new approach for functional and
  morphological imaging,'' \emph{Nature medicine}, vol.~14, no.~4, pp.
  459--465, 2008.

\bibitem{wang2015vision}
G.~Wang, M.~Kalra, V.~Murugan, Y.~Xi, L.~Gjesteby, M.~Getzin, Q.~Yang, W.~Cong,
  and M.~Vannier, ``Vision 20/20: Simultaneous ct-mri—next chapter of
  multimodality imaging,'' \emph{Medical physics}, vol.~42, no.~10, pp.
  5879--5889, 2015.

\bibitem{fahrig2001truly}
R.~Fahrig, K.~Butts, J.~A. Rowlands, R.~Saunders, J.~Stanton, G.~M. Stevens,
  B.~L. Daniel, Z.~Wen, D.~L. Ergun, and N.~J. Pelc, ``A truly hybrid
  interventional mr/x-ray system: Feasibility demonstration,'' \emph{Journal of
  Magnetic Resonance Imaging: An Official Journal of the International Society
  for Magnetic Resonance in Medicine}, vol.~13, no.~2, pp. 294--300, 2001.

\bibitem{fahrig2005performance}
R.~Fahrig, Z.~Wen, A.~Ganguly, G.~DeCrescenzo, J.~Rowlands, G.~Stevens,
  R.~Saunders, and N.~Pelc, ``Performance of a static-anode/flat-panel x-ray
  fluoroscopy system in a diagnostic strength magnetic field: A truly hybrid
  x-ray/mr imaging system,'' \emph{Medical physics}, vol.~32, no. 6Part1, pp.
  1775--1784, 2005.

\bibitem{lagendijk2014magnetic}
J.~J. Lagendijk, B.~W. Raaymakers, and M.~Van~Vulpen, ``The magnetic resonance
  imaging--linac system,'' in \emph{Seminars in radiation oncology}, vol.~24,
  no.~3.\hskip 1em plus 0.5em minus 0.4em\relax Elsevier, 2014, pp. 207--209.

\bibitem{raaymakers2017first}
B.~Raaymakers, I.~J{\"u}rgenliemk-Schulz, G.~Bol, M.~Glitzner, A.~Kotte,
  B.~Van~Asselen, J.~De~Boer, J.~Bluemink, S.~Hackett, M.~Moerland
  \emph{et~al.}, ``First patients treated with a 1.5 t mri-linac: clinical
  proof of concept of a high-precision, high-field mri guided radiotherapy
  treatment,'' \emph{Physics in Medicine \& Biology}, vol.~62, no.~23, p. L41,
  2017.

\bibitem{marques2019low}
J.~P. Marques, F.~F. Simonis, and A.~G. Webb, ``Low-field mri: An mr physics
  perspective,'' \emph{Journal of magnetic resonance imaging}, vol.~49, no.~6,
  pp. 1528--1542, 2019.

\bibitem{mcdaniel2019mr}
P.~C. McDaniel, C.~Z. Cooley, J.~P. Stockmann, and L.~L. Wald, ``The mr cap: A
  single-sided mri system designed for potential point-of-care limited
  field-of-view brain imaging,'' \emph{Magnetic resonance in medicine},
  vol.~82, no.~5, pp. 1946--1960, 2019.

\bibitem{cooley2021portable}
C.~Z. Cooley, P.~C. McDaniel, J.~P. Stockmann, S.~A. Srinivas, S.~F. Cauley,
  M.~{\'S}liwiak, C.~R. Sappo, C.~F. Vaughn, B.~Guerin, M.~S. Rosen
  \emph{et~al.}, ``A portable scanner for magnetic resonance imaging of the
  brain,'' \emph{Nature biomedical engineering}, vol.~5, no.~3, pp. 229--239,
  2021.

\bibitem{liu2021low}
Y.~Liu, A.~T. Leong, Y.~Zhao, L.~Xiao, H.~K. Mak, A.~C.~O. Tsang, G.~K. Lau,
  G.~K. Leung, and E.~X. Wu, ``A low-cost and shielding-free ultra-low-field
  brain mri scanner,'' \emph{Nature communications}, vol.~12, no.~1, pp. 1--14,
  2021.

\bibitem{turner1993gradient}
R.~Turner, ``Gradient coil design: a review of methods,'' \emph{Magnetic
  resonance imaging}, vol.~11, no.~7, pp. 903--920, 1993.

\bibitem{wen2005robust}
Z.~Wen, R.~Fahrig, and N.~J. Pelc, ``Robust x-ray tubes for use within magnetic
  fields of mr scanners,'' \emph{Medical physics}, vol.~32, no. 7Part1, pp.
  2327--2336, 2005.

\bibitem{wang2020deep}
G.~Wang, J.~C. Ye, and B.~De~Man, ``Deep learning for tomographic image
  reconstruction,'' \emph{Nature Machine Intelligence}, vol.~2, no.~12, pp.
  737--748, 2020.

\bibitem{shen2020introduction}
C.~Shen, D.~Nguyen, Z.~Zhou, S.~B. Jiang, B.~Dong, and X.~Jia, ``An
  introduction to deep learning in medical physics: advantages, potential, and
  challenges,'' \emph{Physics in Medicine \& Biology}, vol.~65, no.~5, p.
  05TR01, 2020.

\bibitem{knoll2016joint}
F.~Knoll, M.~Holler, T.~Koesters, R.~Otazo, K.~Bredies, and D.~K. Sodickson,
  ``Joint mr-pet reconstruction using a multi-channel image regularizer,''
  \emph{IEEE transactions on medical imaging}, vol.~36, no.~1, pp. 1--16, 2016.

\bibitem{macovski1996noise}
A.~Macovski, ``Noise in mri,'' \emph{Magnetic resonance in medicine}, vol.~36,
  no.~3, pp. 494--497, 1996.

\bibitem{maggioni2012nonlocal}
M.~Maggioni, V.~Katkovnik, K.~Egiazarian, and A.~Foi, ``Nonlocal
  transform-domain filter for volumetric data denoising and reconstruction,''
  \emph{IEEE transactions on image processing}, vol.~22, no.~1, pp. 119--133,
  2012.

\bibitem{feldkamp1984practical}
L.~A. Feldkamp, L.~C. Davis, and J.~W. Kress, ``Practical cone-beam
  algorithm,'' \emph{Josa a}, vol.~1, no.~6, pp. 612--619, 1984.

\bibitem{yu2009compressed}
H.~Yu and G.~Wang, ``Compressed sensing based interior tomography,''
  \emph{Physics in medicine \& biology}, vol.~54, no.~9, p. 2791, 2009.

\bibitem{van2015astra}
W.~Van~Aarle, W.~J. Palenstijn, J.~De~Beenhouwer, T.~Altantzis, S.~Bals, K.~J.
  Batenburg, and J.~Sijbers, ``The astra toolbox: A platform for advanced
  algorithm development in electron tomography,'' \emph{Ultramicroscopy}, vol.
  157, pp. 35--47, 2015.

\bibitem{shepp1974fourier}
L.~A. Shepp and B.~F. Logan, ``The fourier reconstruction of a head section,''
  \emph{IEEE Transactions on nuclear science}, vol.~21, no.~3, pp. 21--43,
  1974.

\bibitem{gach20082d}
H.~M. Gach, C.~Tanase, and F.~Boada, ``2d \& 3d shepp-logan phantom standards
  for mri,'' in \emph{2008 19th International Conference on Systems
  Engineering}.\hskip 1em plus 0.5em minus 0.4em\relax IEEE, 2008, pp.
  521--526.

\bibitem{ackerman1998visible}
M.~J. Ackerman, ``The visible human project,'' \emph{Proceedings of the IEEE},
  vol.~86, no.~3, pp. 504--511, 1998.

\end{thebibliography}

\end{document}